\documentclass[journal=ancac3,manuscript=article]{achemso}

\usepackage{graphicx}
\usepackage[version=3]{mhchem} 
\usepackage{epstopdf}

\newcommand{\Fig}[1]{Fig.\ref{#1}}
\newcommand{\Figure}[1]{Figure~\ref{#1}}
\newcommand{\SI}[1] {Supporting Information}

\author{Enrique Burzur\'{\i}}
\email{E.BurzuriLinares@tudelft.nl}
\author{Joshua O. Island}
\affiliation{Kavli Institute of Nanoscience, Delft University of Technology, Lorentzweg 1, 2628 CJ Delft, The Netherlands}
\altaffiliation{These two authors contributed equally to this work}
\author{Ra\'{u}l D\'{\i}az-Torres}
\affiliation{Departament de Qu\'{\i}mica Inorg\`{a}nica, Universitat de Barcelona, Diagonal 645, 08028 Barcelona, Spain}
\alsoaffiliation{CSIC-ICMAB (Institut de Ci\`{e}ncia dels Materials de Barcelona) Campus de la Universitat Aut\`{o}noma de Barcelona, 08193 Bellaterra, Spain}
\altaffiliation{These two authors contributed equally to this work}
\author{Alexandra Fursina}
\affiliation{Kavli Institute of Nanoscience, Delft University of Technology, Lorentzweg 1, 2628 CJ Delft, The Netherlands}
\author{Ar\'{a}ntzazu Gonz\'{a}lez-Campo}
\affiliation{CSIC-ICMAB (Institut de Ci\`{e}ncia dels Materials de Barcelona) Campus de la Universitat Aut\`{o}noma de Barcelona, 08193 Bellaterra, Spain}
\author{Olivier Roubeau}
\affiliation{Instituto de Ciencia de Materiales de Arag\'{o}n (ICMA), CSIC and Universidad de Zaragoza, Plaza San Francisco s/n 50009 Zaragoza, Spain}
\author{Simon J. Teat}
\affiliation{Advanced Light Source, Lawrence Berkeley National Laboratory, Berkeley, California 94720, USA}
\author{N\'{u}ria Aliaga-Alcalde}
\email{naliaga@icmab.es}
\affiliation{ICREA (Instituci\'{o}  Catalana de Recerca i Estudis Avan\c{c}ats), Passeig Llu\'{\i}s Companys 23, 08010 Barcelona, Spain}
\alsoaffiliation{CSIC-ICMAB (Institut de Ci\`{e}ncia dels Materials de Barcelona) Campus de la Universitat Aut\`{o}noma de Barcelona, 08193 Bellaterra, Spain}
\author{Eliseo Ruiz}
\affiliation{Departament de Qu\'{\i}mica Inorg\`{a}nica, Universitat de Barcelona, Diagonal 645, 08028 Barcelona, Spain}
\alsoaffiliation{Institut de Recerca de Qu\'{\i}mica Te\`{o}rica i Computacional, Universitat de Barcelona, Diagonal 645, 08028 Barcelona Spain}

\author{Herre S. J. van der Zant}
\affiliation{Kavli Institute of Nanoscience, Delft University of Technology, Lorentzweg 1, 2628 CJ Delft, The Netherlands}

\date{\today}

\title{Sequential Electron Transport and Vibrational Excitations in an Organic Molecule Coupled to Few-Layer Graphene Electrodes}


\begin{document}
\clearpage
\begin{abstract}

Graphene electrodes are promising candidates to improve reproducibility and stability in molecular electronics through new electrode-molecule anchoring strategies. Here we report sequential electron transport in few-layer graphene transistors containing individual curcuminoid-based molecules anchored to the electrodes \textit{via} $\pi-\pi$ orbital bonding. We show the coexistence of inelastic co-tunneling excitations with single-electron transport physics owing to an intermediate molecule-electrode coupling; we argue that an intermediate electron-phonon coupling is the origin of these vibrational-assisted excitations. These experimental observations are complemented with density functional theory calculations to model electron transport and the interaction between electrons and vibrational modes of the curcuminoid molecule. We find that the calculated vibrational modes of the molecule are in agreement with the experimentally observed excitations.
\end{abstract}


Molecular electronics promises to take advantage of the variety of built-in functionalities of single molecules to fabricate molecule-based electronic devices\cite{Aviram1974}. The advance in miniaturization of electronic components has been preceded by a plethora of interesting physics at the nanoscale brought on by the interaction between charge, magnetism and superconductivity at the single molecule limit\cite{Heinrich2015,Island2014}. The progress towards robust, room-temperature operation has, however, been limited by the instability of the molecule-gold bond in ambient conditions inherent to the high mobility of gold atoms\cite{Prins2009,Strachan2006}. Additionally, reproducibility of the molecular conductance remains an open challenge in solid-state devices due to the variability in the geometry of the molecule-electrode bonding from device to device.

Carbon-based electrodes\cite{Guo2006,Marquardt2010} and in particular graphene electrodes\cite{Prins2011,Mol2015,Ullmann2015,Puczkarski2015} are attracting special interest as a viable solution for stability and reproducibility in contacting single molecules. The covalent sp$^2$ hybridization of the carbon lattice provides graphene electrodes with high stability even in ambient conditions\cite{Burzuri2012}. Moreover, the two-dimensional character of graphene reduces the thickness of the electrodes and may therefore increase the coupling of the molecule to an underlying gate. Increasing in complexity, graphene electrodes can be combined with superconducting or ferromagnetic metals to achieve functional hybrid devices. Such geometry preserves the anchoring chemistry between molecule and electrodes\cite{Island2014}.

When using graphene, new anchoring strategies need to be developed to substitute the traditional anchoring groups to gold (-SH, -NH). Graphene offers the possibility of covalent bonds to the molecule leading to stronger bonds stable for room temperature application\cite{Prins2011}. In addition, aromatic anchoring groups such as pyrene or anthracene could lead to a more reproducible electrode-molecule bonding thanks to a gentler contact with the electrode that preserves a more well-defined geometry\cite{Garcia-Suarez2013,Mol2015}. However, the potential of $\pi-\pi$ stacked molecules can be limited by the variety of shapes and composition of the edges of the electrodes\cite{Carrascal2012}. The molecule-electrode contact should therefore be placed far from the edge to minimize its influence.

In this work we synthesize a new curcuminoid-based molecule better suited for few-layer graphene (FLG) junctions improving on our previous study\cite{Prins2011} by extending the transport backbone which allows the anchoring groups to sit farther from the graphene edges. We report intermediate electrode-molecule coupling ($\Gamma$) by $\pi-\pi$ stacking of anthracene groups to FLG electrodes. We show the coexistence of inelastic co-tunneling excitations with Coulomb blockade physics. We further show that an intermediate electron-phonon coupling results in vibrational excitations in the single-electron transport (SET) and Coulomb blockade regimes. To complement our experimental observations, we perform density functional theory (DFT) calculations to model electron transport and the interaction between electrons and vibrational modes of the curcuminoid molecule.

\begin{figure}
\includegraphics[width=0.7\textwidth]{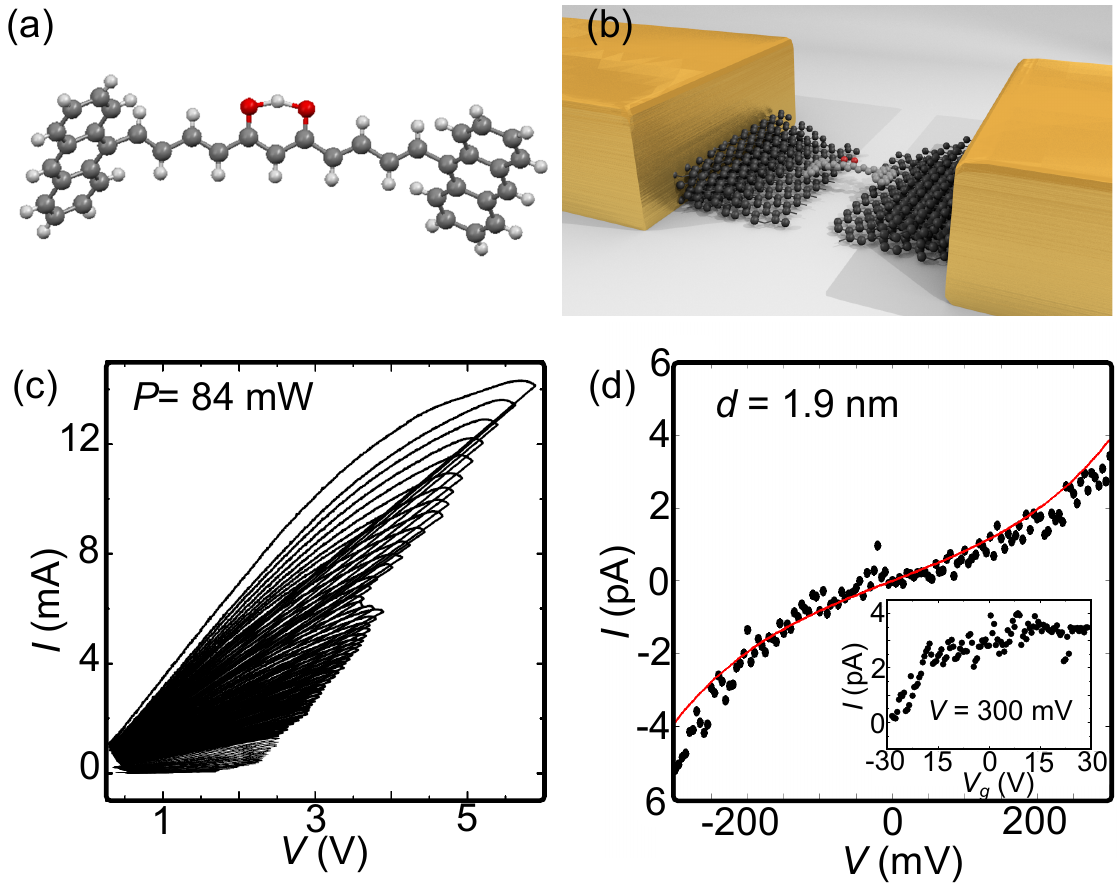}
\caption{(a) Schematics of the curcuminoid-based molecule: 9ALCccmoid. The central curcuminoid backbone is completed with a pair of anthracene groups. Color code: C = grey, O = red, H = white. (b) Artistic representation of a FLG-curcuminoid junction. (c) Typical evolution of the $I-V$ characteristics during the electroburning of a device. (d) Current-voltage trace of an open gap at liquid helium temperature before molecular deposition. The solid line is a fit to the Simmons model with gap size $d=1.9\pm 0.3$ nm, barrier height $\phi\approx 0.63$ eV and gap area $A\approx 5$ nm$^2$. The inset shows the gate trace at $V=300$ mV for the same device. No gate dependence is observed in the accessible gate voltage range.}
\label{figure1}
\end{figure}

\section{\normalsize{RESULTS AND DISCUSSION}}

The structure of the curcuminoid-based molecule, 9ALCccmoid, is schematically shown in \Fig{figure1}(a). This molecule is synthesized by modifying the methodology reported in the literature \cite{Pabon2010,Babu1994,John2006}. Details of the synthesis and crystallographic information are given in the Methods section. The 9ALCccmoid molecule is made of a conjugated linear chain of eleven C atoms that contains a $\beta$-diketone group at the center of the backbone (diarylundecanoid framework). Two anthracene groups complete the chain on both sides. The crystallographic data shows that the conjugated linear backbone presents a gentle bend (see Fig.S2 in the \SI~). The two anthracene arms are tilted from the chain showing a relative torsion angle close to $50.5^{\circ}$ as shown in Fig.S3 in the \SI~. The single molecule does not present symmetry; this way, the $\beta$-diketone moiety, shows two C-O distances, C2-O1 and C22-O2, of 1.300 and 1.290~\AA, respectively; with a calculated intramolecular O1$\cdot\cdot\cdot$H$\cdot\cdot\cdot$O2 angle of 154$^\circ$ and two O$\cdot\cdot\cdot$H theoretical distances, O1$\cdot\cdot\cdot$H of 1.256~\AA~and O2$\cdot\cdot\cdot$H of 1.296~\AA.\footnote{The two possible tautomers have not been included in the explanation of the transport through the curcuminoid. Only the keto/enol tautomer has been used as is indicated in Figs 1 and 5.
Interestingly, the crystallographic structure shows that the keto/enol tautomer is the most stable (Fig 1).  In addition, DFT calculations with the B3LYP functional indicate that for the isolated molecule, the keto/enol form is 15.4 kcal/mol more stable than the diketo one (considering the C2v isomer). This calculated energy difference is even higher than the equivalent for the acetylacetone (acac) molecule (10.0 kcal/mol \cite{Alagona2008}). Thus, the estimated interconversion barrier for acac is higher than 60 kcal/mol and should be even higher for the curcuminoid confirming the stability of the keto/enol form due to the strong hydrogen bond.
}The conjugated $\pi$ system is well-characterized with alternating C-C and C=C distances having values between 1.431-1.470 (C2-C3, C4-C5, C6-C7, C22-C23, C24-C25 and C26-C27) and 1.334-1.399~\AA (C1-C2, C3-C4, C5-C6, C1-C22, C23-C24 and C25-C26), correspondingly. Table~\ref{table1} summarizes some selected interatomic distances.

\begin{table}[htb]
\begin{center}
\begin{tabular}{lllllll}
\hline
\textbf{O1 C2}& & 1.300(2) && \textbf{C2 O1 H1} && 102.5(10)\\
\textbf{O1 H1}& & 1.24(2) & &\textbf{C22 O2 H1}&& 101.3(9)\\
\textbf{O2 C22}& & 1.291(2) && \textbf{C2 C1 C22}&& 120.57(19)\\
\textbf{O2 H1}& & 1.31(2) && \textbf{O1 C2 C1} && 120.54(18)\\
\textbf{C1 C2}& & 1.395(3) && \textbf{O1 C2 C3} && 117.94(18)\\
\textbf{C1 C22}& & 1.397(3) && \textbf{C1 C2 C3}&& 121.48(19)\\
\textbf{C2 C3}& & 1.452(3) && \textbf{C4 C3 C2} && 123.20(19)\\
\textbf{C3 C4}& & 1.341(3) && \textbf{C3 C4 C5} && 125.15(19)\\
\textbf{C4 C5}& & 1.438(3) && \textbf{C6 C5 C4} && 121.88(19)\\
\textbf{C6 C7}& & 1.466(3) && \textbf{C5 C6 C7} && 127.84(19)\\
\hline
\end{tabular}
\caption{Selected interatomic distances (\AA) and angles (deg) for 9ALCccmoid.}
\label{table1}
\end{center}
\end{table}

The length of the molecule is 2.15 nm and therefore is significantly longer than the antecedent ccmoid form 9Accm\cite{Menelaou2013,Aliaga-Alcalde2012,Menelaou2012,Aliaga-Alcalde2010}, with a length of 1.71 nm and only seven C atoms in the chain \cite{Prins2011}. This modification ideally brings the anchoring groups farther from the edges, facilitating the coupling of the molecule to the aromatic structure of the graphene by $\pi-\pi$ stacking. The $\beta$-diketone moiety and the two ending anthracene groups are identical for both systems. The electroburning process is known to result in a variety of source-drain separations ranging from 1-2 nm\cite{Prins2011, Burzuri2012}. The purpose behind the design of 9ALCccmoid is to provide an optimal length, capable of adapting the molecule to a range of electrode separations without disturbing the conjugation of the molecule and the anchoring groups. The comparison of the two structures 9ALCccmod and 9Accm, further characterization of 9ALCccmoid and the resulting lengths of both molecules are shown in the \SI~.

The nanometer-spaced electrodes are fabricated by electroburning of few-layer graphene (FLG) flakes contacted by gold leads as sketched in \Fig{figure1}(b). Flakes of around 10 nm thickness are obtained by mechanical exfoliation of natural graphite and deposited onto a silicon wafer coated with 285 nm of silicon oxide. The underlying silicon substrate is used as gate electrode. Further details of the fabrication of the devices are described elsewhere\cite{Prins2011,Burzuri2012,Island2014}. Few-layer graphene flakes are selected, in contrast to single layer, to guarantee a continuous reservoir of electrons and avoid the gating of the source and drain electrodes \cite{Datta2009,Castellanos-Gomez2012}. The gating of the current in our devices can therefore be ascribed to the gating of the molecule itself. This is an improvement when compared with previous reports on carbon-based electrodes such as carbon nanotubes\cite{Qi2004,Guo2006}. \Figure{figure1}(c) shows the step-by-step drop of the current ($I$) during the controlled electroburning of a typical device. Initial powers around 80 mW and temperatures of 1000 K are required to start the burning process \cite{Island2014}. The molecules are therefore deposited after the electroburning to preserve their integrity.

Prior to molecule deposition, the devices with the open gaps are further characterized in vacuum and at liquid helium temperatures. \Figure{figure1}(d) shows the $I-V$ characteristic of a typical device. The s-shape exponential dependence of the measured source-drain current with applied bias voltage, $V$, is a fingerprint of a tunneling current between the electrodes. The solid line is a fit using the Simmons model\cite{Simmons1963,Prins2011} for a current through a thin insulator. The size of the gap obtained from the fit is $d=1.9\pm0.3$ nm. Other fitting parameters are the height of the barrier ($\phi\approx$ 0.63 eV) and the area of the gap ($A\approx$ 5 nm$^2$). The low barrier value compared with reported values for graphene (4.5 eV) is a common occurrence when the Simmons' model is applied to I-V characteristics of nanoscale junctions.\cite{Mangin2009,Wang2011,Island2014}. The gate trace corresponding to the same device is measured at $V=300$ mV. Except for a slow capacitive increase in current due to the high junction resistance, no gate dependence is observed in the accessible gate voltage window ($-40< V_g <40$ V) as seen in the inset of \Fig{figure1}(d). This is a strong indication of an empty gap. Note that such a gate response at low temperatures stipulates the absence of small graphene quantum dots in the gap as room-temperature characterization may overlook graphene nano-islands with large charging energies that could mimic the behaviour of the target molecules\cite{Barreiro2012} at low temperatures.

Twenty three FLG flakes have been electroburned and characterized at liquid helium temperatures. The curcuminoid molecules are afterwards deposited at room temperature by drop-casting of a 0.1 mM solution of the synthesized molecules in dichloromethane. Two of these devices showed an increase of the current after molecule deposition. This yield is typical for junctions fabricated by electroburning\cite{Prins2011,Burzuri2012}. Importantly, high conductance resonances could be resolved in the gate trace for one of the samples and a second sample shows inelastic co-tunneling excitations of similar energy independent of the gate voltage. The charge transport measurements are performed at liquid helium temperatures and in vacuum.

\begin{figure}
\includegraphics[width=0.7\textwidth]{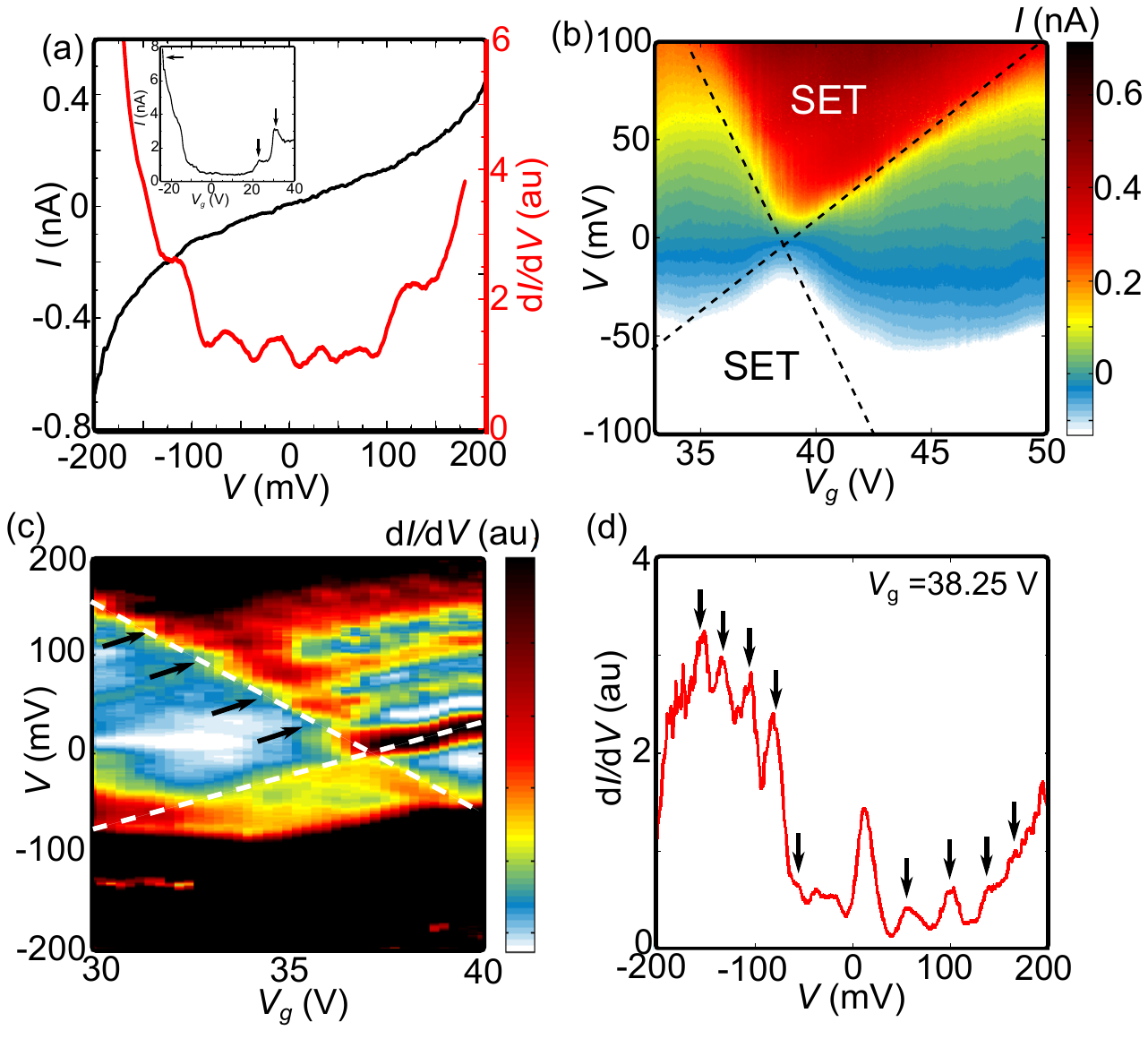}
\caption{(a) Current-voltage ($I-V$) and differential conductance (d$I$/d$V$) traces measured in sample A at $T=4$ K and $V_g$=0 after molecule deposition at room $T$. The current increases two orders of magnitude compared to the empty gap in \Fig{figure1}(d). Several steps appear in $I$ and the corresponding peaks in d$I$/d$V$ . [Inset] $I-V_g$ trace at $T=4$ K and $V=200$ mV. Three high-current peaks appear at 23 V, 31 V and below -20 V. (b) Current color plot measured in sample A as a function of $V$ and $V_g$ for positive $V_g$ values. High current regions, a signature of sequential electron transport (SET), separate low-current regions where current is blocked. The dashed lines mark the edges of the diamonds. (c) $\textrm{d}I/\textrm{d}V$ color plot numerically derived from (b) in sample A. Several SET excitations (marked with black arrows) appear at positive bias at 46($\pm$6), 79($\pm$6), 115($\pm$10) and 150($\pm$10) mV. (d) d$I$/d$V$ trace measured on sample A at $V_g=38.25$ V in which the excitations are also visible symmetric in bias. The black arrows mark the position of the excitations at this specific $V_g$. The dashed lines mark the edges of the diamonds.}
\label{figure2}
\end{figure}

\Figure{figure2}(a) shows the $I-V$ and d$I$/d$V$ traces after molecule deposition on the same device described in \Fig{figure1}, hereafter sample A. The d$I$/d$V$ is obtained by numerical differentiation of $I$. After molecule deposition, the current increases around two orders of magnitude and becomes asymmetric in bias. Moreover, several steps appear in the current that are more clearly seen as peaks in the d$I$/d$V$. The current level is comparable with that reported for the shorter curcuminoid derivative \cite{Prins2011}. This picture indicates that either the anchoring is the limiting factor for the current in this family of molecules or that the effective transport pathway of the molecules is similar. Our theoretical results (see below) support the transport pathway picture.

The inset of \Fig{figure2}(a) shows the $I-V_g$ trace measured at $V = 200$ mV. Two high-current peaks appear at 23 V and 31 V in contrast with the flat $I-V_g$ for the empty junction in \Fig{figure1}(d). An additional resonance starts to show up below -20 V. These peaks are a fingerprint of resonant transport through a molecule coupled to the electrodes. \Figure{figure2}(b) shows an $I$ color plot measured in sample A as a function of $V$ and $V_g$ around the $V_g=31$ V resonance (for a complete color plot in $V_g$ see \SI~). High-current regions, a signature of sequential electron transport (SET), separate low-current regions where current is blocked. In the later, the charge in the molecule is stabilized; \textit{i.e.}, the molecule is in the Coulomb blockade regime. \Figure{figure2}(c) shows a d$I$/d$V$ color plot numerically derived from a finer $I$ measurement around the SET region shown in \Fig{figure2}(b). High-conductance SET resonances separate low-conductance regions. Additional inelastic co-tunneling excitations within the low-conductance diamonds (not visible in this contrast) point to an intermediate coupling of the molecule with the electrodes ($\Gamma$)\cite{Perrin2015}. The value of $\Gamma$ is estimated from the full width at half maximum (FWHM) of the d$I$/d$V$ peaks at the Coulomb diamond edge. We find $\Gamma=10$ meV (see \SI~ for calculation), a value that is comparable to that obtained with thiol groups used for gold electrodes\cite{Perrin2015}. This result shows that an intermediate $\Gamma$ can be achieved with $\pi-\pi$ stacked molecules on FLG electrodes. The gate coupling of the molecule is calculated to be $\beta=0.006$ (see \SI~ for calculation) which allows a shift of the molecular levels of around 480 meV in our devices.

Interestingly, high-bias SET lines running parallel to the edges of the diamonds are observed at 46($\pm$6), 79($\pm$6), 115($\pm$10) and 150($\pm$10) mV at positive bias. The energy of the excitations is taken from the crossing of the lines with the opposite diamond edge. Hints of an additional excitation are observed at around 20 mV. However, the broadening of the Coulomb edge induced by $\Gamma$ hinders the resolution of individual levels below 20-25 mV. Negative bias excitations that end in the adjacent charge state are only clearly seen for $-62$($\pm$6) mV. For clarity, \Figure{figure2}(d) shows a d$I$/d$V$ trace measured at $V_g=38.25$ V where the excitations are signaled by black arrows. Here, negative bias excitations become also visible. These excitations are masked in \Fig{figure2}(c) due to the low color contrast from higher current at negative bias. The positive-bias peak close to $V = 0$ is the edge of the diamond.

\begin{figure}
\includegraphics[width=0.4\textwidth]{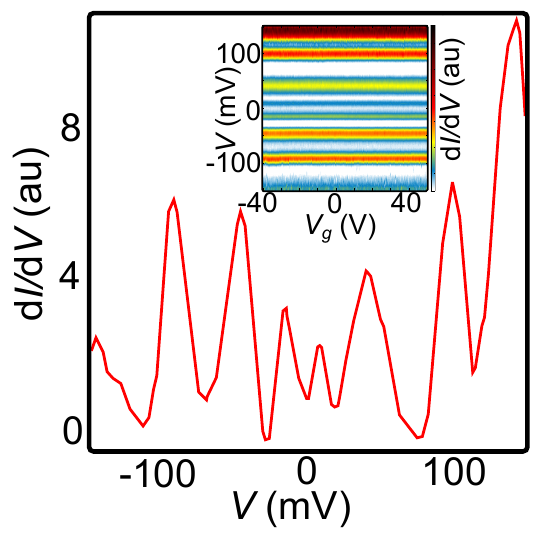}
\caption{d$I$/d$V$ trace measured at $V_g=0$ in sample B. Several excitations independent of $V_g$ appear symmetric in bias at -144, -91, -46, -16, 8, 42, 99 and 142 mV. The inset shows a d$I$/d$V$ color plot measured in sample B as a function of $V$ and $V_g$ at $T=2$ K.  No resonant transport is observed.}
\label{figure3}
\end{figure}

The color d$I$/d$V$ map of an additional junction, hereafter sample B, is shown in the inset of \Fig{figure3}. Several inelastic co-tunneling excitations appear symmetric in bias that are independent of the gate voltage. No resonant transport is observed in the gate range in contrast to the sample A. The absence of resonant transport could be ascribed to a lower gate coupling that shifts the resonances to higher absolute gate voltages. A representative d$I$/d$V$ trace at $V_g=0$ is shown in \Fig{figure3}. Interestingly, the energies of the excitations: -144($\pm$2), -91($\pm$2), -46($\pm$2), 42($\pm$2), 99($\pm$2) and 142($\pm$2) mV approximately agree with those observed in the sample A at positive bias. In addition, sample B shows two low-bias additional excitations at -16($\pm$2) mV and 8($\pm$2) mV. These new resonances could be masked in sample A by the broadening induced by $\Gamma$ in resonant transport. Table~\ref{table2} lists the energy of the excitations in both samples. The origin of such excitations in 9ALCccmoid could be either magnetic, electronic or vibrational. The 9ALCccmoid is a non-magnetic molecule and therefore we disregard magnetic excitations as in Ref.\cite{Wagner2013}. We argue below in favor of the vibrational origin of the excitations due to an intermediate electron-phonon coupling.

\begin{figure}
\includegraphics[width=1\textwidth]{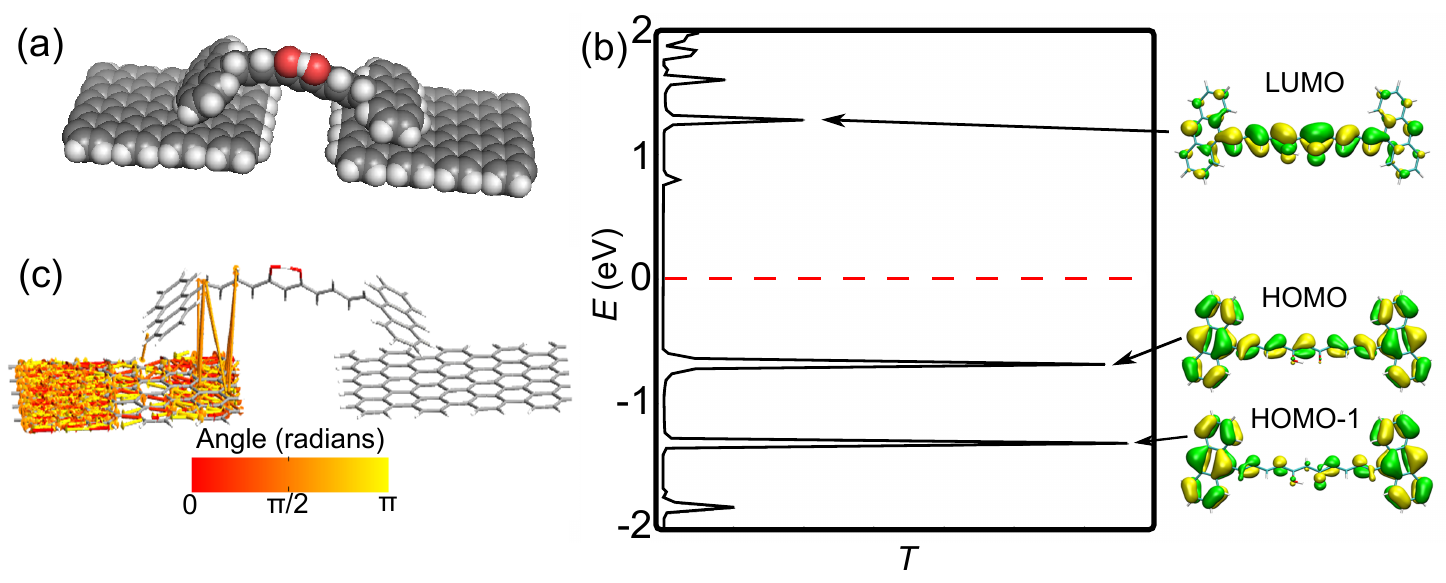}
\caption{(a) Optimized geometry of the 9ALCccmoid molecule on top of graphene electrodes. (b) Transmission spectrum of the 9ALCccmoid curcuminoid and corresponding iso-surfaces for the HOMO, HOMO-1 and LUMO orbitals. (c) Transmission pathway of the 9ALCccmoid on graphene electrodes. The color scale is the direction of the transport in radians: 0 is for rightwards transport , $\pi$ is for leftwards transport and $\pi /2$ is transport perpendicular to the graphene plane. The calculations show that electrons jump on the molecule through the anthracene but also through the backbone.}
\label{figure4}
\end{figure}

To gain insight on these experimental observations, we have performed DFT calculations on the curcuminoid molecule. DFT calculations of the molecular device were performed with the numerical ATK 2014.2 code\cite{QuantumWiseA/S2014,Brandbyge2002,Soler2002} using the PBE functional adding dispersion terms through Grimme D2 approach\cite{Grimme2006}. The model structure was created from a single layer of graphene by deleting a ribbon of four carbon atoms and adding hydrogen atoms to saturate the carbon dangling bonds. The 9ALCccmoid molecule was placed on top with one anthracene group placed close to each graphene electrode. The fully optimized structure is shown in \Fig{figure4}(a). The non-symmetric final structure is due to the presence of an energy minimum with a small relative horizontal shift of the graphene electrodes and it was selected in order to be closer to the experimental geometry.

\Figure{figure4}(b) shows the transmission through the molecule as a function of the energy. The energy of the HOMO-LUMO gap is approximately 2 eV. In this non-magnetic molecule, the first excited electronic levels are expected to appear close to the energy of the gap, that is, two orders of magnitude higher than the energies of the observed excitations. Therefore the origin of the excited lines is not expected to be due to electronic excitations. The energy isosurfaces of the HOMO, HOMO-1 and LUMO orbitals are shown associated with their respective transmission peaks. The LUMO is delocalized over the backbone of the molecule and the $\beta$-diketone group. The HOMO and HOMO-1 are also strongly present in the anthracene anchoring groups favoring the current flow through the molecule. This last aspect clearly manifests itself in the higher transmission of the HOMO and HOMO-1. In contrast, the HOMO is only weakly present in the $\beta$-diketone group. Interestingly, the H atom in the $\beta$-diketone group can be replaced by different magnetic metallic atoms\cite{Menelaou2013,Aliaga-Alcalde2012,Menelaou2012,Aliaga-Alcalde2010}. The small presence of the $\beta$-diketone for HOMO transport makes 9ALCccmoid a good candidate to probe magnetism at the nanoscale by providing a well conjugated backbone in close proximity to a magnetic atom.

\Figure{figure4}(c) shows the calculated transmission pathway for the 9ALCccmoid molecule on graphene electrodes. The calculations assume that the electrons are injected from the left electrode. The transmission thereafter fades through the molecule and is below the threshold set by the colour scale when it reaches the right-side electrode. The calculations suggest that the electrons tunnel through the molecule not only \textit{via} the anthracene group but strongly \textit{via} the backbone closer to the gap, even if the backbone is farther from the graphene. The explanation can be found by inspecting the orbitals in \Fig{figure4}(b). The LUMO, the most active for transport, is mainly centered in the backbone and barely present in the anthracene groups. This picture could explain the relatively similar currents obtained for the 9ALCccmoid and the shorter 9ALCccmod curcuminoid molecules, because in both cases the transport path is essentially determined by the distance between the graphene source and drain electrodes.

\begin{table}[htb]
\begin{center}
\begin{tabular}{|l|l|l|l|}
\hline
$\sharp$ & Sample A. \textit{E} (meV) & Sample B. \textit{E} (meV) & Theory. \textit{E} (meV)\\
\hline \hline
1 & 150 & 142 & 145\\
2 & 115 & X & X\\
3 & 79 & 99 & 89.6 \\
4 & 46 & 42 & 48.5 \\
5 & X & 8 & 6-14* \\
\hline
\end{tabular}
\caption{Energy of the SET excitations for sample A, inelastic co-tunneling excitations for sample B and energy of the calculated excitations with intermediate Franck-Condon coupling.* The calculations predict several modes between 6 and 14 meV with high $\lambda$.}
\label{table2}
\end{center}
\end{table}

To understand the origin of the excitations observed in the resonant transport regime in sample A and the Coulomb blockade regime in sample B we calculate the vibrational modes for the neutral, anion and cation curcuminoid molecule without electrodes. This is justified because the molecule is coupled by van der Waals interactions to the electrodes\cite{Burzuri2014}. Electron-phonon coupling of the isolated 9ALCccmoid molecule was calculated by using the Franck-Condon module of the Gaussian09 (release d01) code\cite{Becke1993} using the B3LYP functional\cite{Frisch2009} and a $6-311+G**$ basis set. The calculations were performed assuming that the neutral molecule gains one electron as that is the most likely scenario: according to our calculations the free energies for the cation and anion with respect to the neutral molecule are 6270 and 2106 meV respectively. The electron-phonon coupling $\lambda$ parameter was determined from the Duschinsky shift vector\cite{Duschinsky1937,Barone2009} $k$ for each calculated vibrational mode with harmonic frequency $\nu$:\cite{Seldenthuis2008}

\begin{equation}
\lambda=k\sqrt{\frac{\pi\nu}{\hbar}}
\end{equation}

\begin{figure}
\includegraphics[width=1\textwidth]{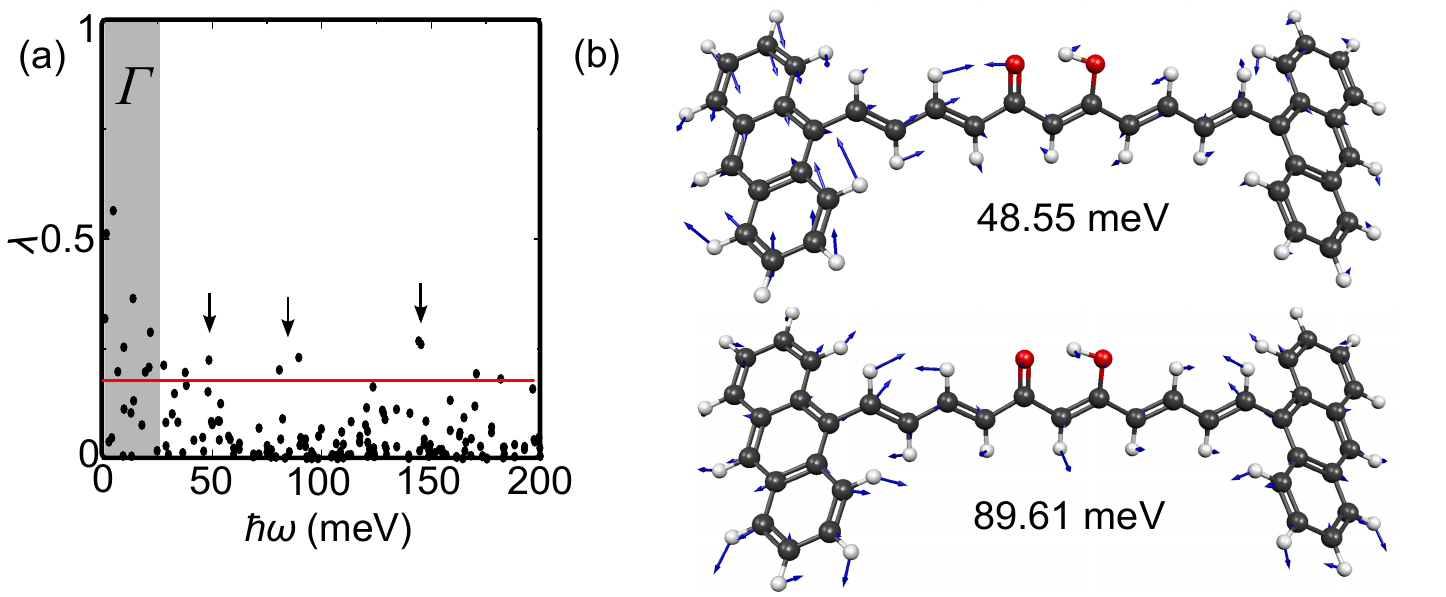}
\caption{(a) Electron-phonon coupling $\lambda$ of the vibrational modes of the molecule. The arrows mark the intermediate $\lambda$ vibrations that match in energy with the experimental observations. The shaded area is the approximate broadening induced by $\Gamma$ on the Coulomb edge. Vibrational modes with lower energy can not be resolved independently. (b) Two representations of vibrational modes with intermediate $\lambda$ and energies 48.55 meV and 89.61 meV. The main contribution to the modes are torsions of the C-H bonds located in the anthracene anchoring groups.}
\label{figure5}
\end{figure}

\Figure{figure5}(a) shows the electron-phonon coupling $\lambda$ of the different vibrational modes of the neutral curcuminoid. Several low-frequency vibrational modes ($\hbar\omega< 25$ meV) show a large $\lambda$ but not large enough to induce Franck-Condon blockade\cite{Burzuri2014}. These excitations can not be resolved independently in \Fig{figure2}(c) due to the broadening of the Coulomb edge induced by $\Gamma$ (shaded area in \Fig{figure5}(a)). These modes may, however, contribute to the higher conductance and broadening of the Coulomb edge when compared with the other edges of the diamond. Interestingly, a few additional vibrational modes with energies around 50, 90 and 145 meV show also an intermediate $\lambda$. In addition, the energy separation between these modes is larger than the estimated broadening induced by $\Gamma$. These energies approximately match with the energies of the SET lines observed in \Fig{figure2}(d) in sample A and in \Fig{figure3} in sample B. We therefore associate the origin of the excitations to vibrational modes of the molecule. The characteristics of the high-energy vibrational modes that have an intermediate $\lambda$ are summarized in two representative modes with energies 48.55 meV and 89.61 meV in \Figure{figure5}(b). The main contributions to the modes are torsions of the C-H bonds located in the anthracene anchoring groups.

\section{\normalsize{CONCLUSIONS}}

In conclusion, a 9ALCccmoid curcuminoid-based molecule is synthesized suited to accommodate the variety of gaps formed from the electromigration of FLG. We demonstrate electrical gating of this long curcuminoid molecule coupled \textit{via} $\pi-\pi$ stacking to FLG electrodes. An intermediate coupling to the electrodes ($\Gamma=10$ meV) can be achieved for $\pi-\pi$ stacked molecules on few-layer graphene electrodes, although our calculations suggest that the molecular backbone can take an active part as a direct transport pathway to the electrodes. Additional high-energy excitations appear running parallel to the Coulomb edges and in the co-tunneling regime. We associate these excitations to vibrations of the curcuminoid due to an intermediate electron-vibron coupling as further supported by our DFT calculations. The modes involved are high-energy modes associated with torsions of the C-H bonds located in the anchoring groups. These results shed light on $\pi-\pi$ bonding strategies in molecular transport studies and show that intermediate coupling can be achieved to observe higher order processes such as co-tunneling and vibrational excitations.

\section{\normalsize{METHODS}}

\subsection{Synthesis of 9ALCccmoid}

All experiments were carried out in aerobic conditions using commercial grade solvents for the synthesis of 9ALCccmoid. The molecule was synthesized according to the procedure described elsewhere\cite{Pabon2010,Babu1994,John2006}. Acetylacetonate (0.83 g, 8.30 mmol) and B$_2$O$_3$ (0.44 g, 6.25 mmol) were dissolved in 10 mL of EtO$_2$CMe. The reaction mixture was heated at $40^\circ$C for 30 min. Then, a solution of 4.1 g of 3-(9-Anthryl)acrolein (17.65 mmol) and 8.12 g of tributyl borate (35.30 mmol) in 20 mL of EtO$_2$CMe was added. The mixture was stirred at $40^\circ$C for 3 h. After cooling down, an excess of n-butylamine (0.44 mL, 4.45 mmol) in EtO$_2$CMe (10 mL) was introduced dropwise. The final reaction was stirred at room temperature for 2 days. As a result, a red solid precipitate appeared. The colored solid was filtered and washed with H$_2$O, MeOH and Et$_2$O to remove impurities. Yield $87~\%$. Crystals of 9ALCccmoid were achieved by slow evaporation using THF. IR data (KBr, cm$^{-1}$) 3424(br), 3068(s), 3041(s), 3028(s), 1603(vs), 1506(s), 1440(s), 1287(s), 1121(vs), 1000(vs), 958 (vs), 884(vs), 842(s), 782(s), 728(vs), 545(s). Elemental analysis calculated for C$_{39}$H$_{28}$O$_{2}\cdot$0.5H$_2$O (537.21 g$\cdot$ mol−1): C 87.11; H 5.44. Found: C 87.01; H 5.37. $^1$H NMR (300 MHz, CDCl$_3$): $\delta$ 8.437 (s, 2H), 8.273 (m, Hantra), 8.024 (m, xH), 7.877 (d, 2H), 7.769 (dd, 2H), 7.503 (m, Hantra), 6.852 (dd, 2H), 6.262 (d, 2H), 5.814 (s, 1H). MALDI+ (m/z): 527.2 [M-H]$^+$.

\subsection{Physical Measurements.}

C and H analyses were performed with a Perkin-Elmer 2400 series II analyzer. Mass spectra were recorded in CHCl$_3$/MeCN (1:1) using matrix assisted laser desorption ionization with time of flight (MALDI-TOF) mass spectrometer (4800 \textit{Plus} MALDI TOF/TOF (ABSciex - 2010)). DHB stands for 2,5-dihydroxybenzoic acid. Infrared spectra (4000-400 cm$^{-1}$) were recorded from KBr pellets on a Bruker IFS-125 FT-IR spectrophotometer.

\subsection{\normalsize{Crystallography}}

The crystalline structure of the 9ALCccmoid derivative has been obtained using synchrotron radiation source. Data were collected on a cut piece of a red needle of dimensions 0.12 x 0.02 x 0.02 mm$^3$ on a Bruker D8 diffractometer on the Advanced Light Source beam-line 11.3.1 at Lawrence Berkeley National Laboratory, from a silicon 111 monochromator ($\lambda$ = 0.7749 ${\AA}$). Data reduction and absorption corrections were performed with SAINT and SADABS\cite{BrukerAXSInc.MadisonWisconsin}, respectively. The structure was solved and refined on F$^2$ with SHELXTL suite\cite{Sheldrick2008,Sheldrick2015,Sheldrick2015a}. Hydrogen atoms were all found in a difference Fourier map, included at calculated positions on their carrier atom and refined with a riding model, except that on the central diketone moiety that was refined freely with its isotropic thermal parameter 1.5 times that of the closest oxygen, O1. The main crystal parameters are summarized in Table~S1 in the \SI~. All crystallographic details can be found in CCDC 1437644 and can be obtained free of charge from the Cambridge Crystallographic Data Centre \textit{via} https://summary.ccdc.cam.ac.uk/structure-summary-form.

\begin{acknowledgement}

This work was supported by the EU FP7 program through project 618082 ACMOL and ERC grant advanced Mols@Mols. It was also supported by OCW and the Dutch funding organization NWO (VENI) and FOM. E.R. thanks Generalitat de Catalunya for an ICREA Academia Award. N.A.-A., R.D.-T. and A.G.-C. thank the MICINN of Spain (projects CTQ2012-32247 and MAT2013-47869-C4-2-P). The Advanced Light Source is supported by the Director, Office of Science, Office of Basic Energy Sciences of the U.S. Department of Energy under contract DE-AC02-05CH11231.

\end{acknowledgement}

\begin{suppinfo}

Contains additional details about the crystalline structure and geometry of the 9ALCccmoid molecule.

\end{suppinfo}

\bibliographystyle{plain}
\bibliography{Curcuminoids}


\end{document}